\documentclass[aps,prl,twocolumn]{revtex4}
\usepackage{subfig}

\usepackage{graphicx}

\def\be{\begin{equation}}
\def\ee{\end{equation}}
\def\bea{\begin{eqnarray}}
\def\eea{\end{eqnarray}}

\begin{document}


\title{Making sense of the bizarre behaviour of horizons 
in the McVittie spacetime}


\author{Valerio Faraoni}
\email[]{vfaraoni@ubishops.ca}
\affiliation{Physics Department and {\em STAR} Research Cluster, 
Bishop's University, 2600 College Street, 
Sherbrooke, Qu\'ebec, Canada J1M~1Z7}

\author{Andres F. Zambrano Moreno}
\email[]{azambrano07@UBishops.ca}
\affiliation{Physics Department,  
Bishop's University, 2600 College Street, 
Sherbrooke, Qu\'ebec, Canada J1M~1Z7}

\author{Roshina Nandra}
\email[]{rn288@mrao.cam.ac.uk}
\affiliation{Astrophysics Group, Cavendish Laboratory, JJ 
Thomson Avenue, Cambridge CB3 0HE, U.K.}
\affiliation{Kavli Institute 
for Cosmology, c/o Institute of Astronomy, Madingley Road, 
Cambridge CB3~0HA, U.K.}



\begin{abstract}
The bizarre behaviour of the apparent (black hole and 
cosmological) horizons of the McVittie spacetime is 
discussed using, as an analogy, the Schwarzschild-de 
Sitter-Kottler spacetime (which is a special case of 
McVittie anyway). For a dust-dominated ``background'' 
universe, a black hole cannot exist at early times because 
its (apparent) horizon would be larger than the 
cosmological (apparent) horizon. A phantom-dominated 
``background'' universe causes this situation, and the 
horizon behaviour, to be time-reversed.  
\end{abstract}


\maketitle



\section{Introduction}

Cosmology and black holes as seen through the eyes of 
general relativity come together in the investigation of a 
dynamical black hole embedded in a cosmological background. 
The interplay between the cosmic dynamics and the black 
hole gives rise to interesting phenomena and can reveal 
some unexpected features of the underlying theory of 
gravity.  In this work we restrict our attention, for 
simplicity, to spherically symmetric systems. The prototypical 
solution of the Einstein equations representing a black 
hole embedded in a cosmological spacetime is the 
Schwarzschild-de Sitter-Kottler solution. This metric is 
special since it admits a timelike 
Killing vector and is, therefore, static in the region 
between the black hole horizon and the de Sitter 
(cosmological) horizon. A less well known solution is the 
1933 McVittie 
solution \cite{McVittie}, which is a generalization of  the 
Schwarzschild-de Sitter-Kottler solution.  In this case the 
black hole is embedded in a  general  
Friedmann-Lema\^itre-Robertson-Walker (FLRW) background, 
so that the region between the black hole horizon and 
cosmological horizon need not be static.  Although it has 
been studied and celebrated by 
many authors \cite{McVittieworks, Nolan, Kleban, Roshina}, 
it has proved  surprisingly difficult to understand (see 
the recent work \cite{AbdelqaderLake}). A simplifying 
assumption in the study of this solution, explicitly 
stated in McVittie's original paper, is the no-accretion
condition $G_0^1=0$ (in spherical coordinates, where 
$G_{\mu\nu}$ is the Einstein tensor). This explicitly 
forbids any radial flow  of material, which should 
otherwise occur whenever a spherically symmetric local 
inhomogeneity (such as a central black  hole) is 
introduced in the background. When this is modelled 
however, more general 
solutions of Einstein's theory become possible. These 
include some generalized McVittie solutions 
\cite{FaraoniJacques, Gaoetal1, Gaoetal2}; solutions such 
as those derived by Husain-Martinez-Nu\~nez  
\cite{HusainMartinezNunez}, Fonarev \cite{Fonarev}, 
Sultana-Dyer \cite{SultanaDyer} and 
McClure-Dyer \cite{McClureDyer}; the class of solutions 
found by Szekeres  \cite{Szekeres, Szafron, Barrow}; 
Lema\^itre-Tolman-Bondi black  hole solutions 
\cite{LTBblackholes}; and other solutions 
\cite{cosmologicalblackholes}. In extended 
theories  of gravity, such as scalar-tensor and $f(R)$ 
gravity, 
several other solutions of the relevant field equations 
(which involve an extra gravitational scalar field or 
higher  derivative terms, respectively) have been found 
and sometimes discussed \cite{Barrow,  CliftonMotaBarrow, 
Clifton, myClifton}. 

The original motivation for McVittie's work \cite{McVittie} was 
the investigation of the effects of the cosmological 
expansion on local systems. Another approach to this 
problem later led to the  construction of Swiss-cheese 
models by Einstein and Straus  \cite{EinsteinStraus}.  
However, although this problem has stimulated much 
discussion over the years \cite{expansionpluslocal}, the 
scientific community as a whole is yet to arrive at an 
agreement about the best approach to it (see 
the recent review \cite{CarreraGiulini}).  When solutions 
representing local inhomogeneities in cosmic backgrounds 
are considered, the scope of the investigation broadens. 
For example, a problem of current interest is the possible 
spatial and temporal variation of the gravitational 
``constant'' (which becomes a scalar field in Brans-Dicke 
and scalar-tensor gravity) \cite{CliftonMotaBarrow}. We  
now know several solutions of this kind, but before 
enlarging the catalog further it is important to fully  
understand the presently known solutions 
(for some of them, it is not even known whether the local 
inhomogeneity is associated with a black hole, a naked 
singularity, or another kind of object). For this reason, 
we revisit here the no-accretion  McVittie solution, 
proposing a quick  way of locating the associated  black 
hole and cosmological (apparent)  horizons and studying 
their evolution. We extend the type of 
cosmological background to include phantom universes, which 
have not been considered before in relation to the McVittie 
solution. 

With the exception of the Schwarzschild-de Sitter-Kottler 
solution, which incorporates only a static background 
universe, spherically 
symmetric black holes in more general  cosmological 
backgrounds are 
dynamical. This significantly complicates their analyses. 
 Since the solutions of Einstein's
equations corresponding to the McVittie metric are highly 
dynamical, it is not convenient for us to study the  event 
horizons 
(both black hole and cosmological), which may not even 
exist. It is more instructive to study the dynamical  
apparent horizons, the importance of which is being 
increasingly recognized in the literature 
\cite{dynamicalhorizons}. It is known that, 
for dynamical cosmological black holes, apparent horizons 
can appear or disappear \cite{Nolan, Kleban, 
FaraoniJacques, Gaoetal1}, and we would like to shed 
some light on this bizarre phenomenology.

The plan of this paper is as follows. In Sec.~II we 
briefly  review the Schwarzschild-de Sitter-Kottler 
solution; this (over-)simplified 
situation will serve us well when attempting to understand 
the more complicated phenomenology of dynamical apparent 
horizons.  In Sec.~III we locate the apparent 
horizons of the McVittie metric for non-phantom 
cosmological backgrounds and recover the previous results 
in certain limits.  
 We then continue with the analysis of 
phantom background universes. Finally, Sec.~IV contains a 
discussion  of our results and our
conclusions. Throughout this work  we use units in which 
the speed of light $c$ 
and Newton's constant $G$ are unity, and we mostly follow 
the notations of Ref.~\cite{Wald}. In particular, the 
metric signature is $-+++$.

\section{The Schwarzschild-de Sitter-Kottler black hole}

The Schwarzschild-de Sitter-Kottler solution  is the 
prototypical  
solution representing a black hole embedded in a 
cosmological background (for a certain range of parameter 
values). We will discuss  the McVittie 
metric by using an analogy with the Schwarzschild-de 
Sitter-Kottler metric wherever possible, even though the 
latter corresponds to a very special situation by 
admitting only a {\em static} black hole in the de Sitter 
background, and its apparent horizons are also event 
horizons. Nonetheless, analogies are made possible by the 
fact  that the 
Schwarzschild-de Sitter-Kottler solution is contained as a  
special case in the  McVittie class of solutions.

The spherically symmetric Schwarzschild-de Sitter-Kottler 
solution of the Einstein equations has line element 
\begin{widetext}
\be \label{1}
ds^2=-\left( 1-\frac{2m}{r}-H^2r^2 \right) dt^2+
\left( 1-\frac{2m}{r}-H^2r^2 \right)^{-1}  dr^2+ r^2 
d\Omega_{(2)}^2 \,,\label{eq:SDSK}
\ee
\end{widetext}
where $r$ is the 
areal radius (of a sphere with surface area $4\pi r^2$),
$ d\Omega_{(2)}^2 =d\theta^2+\sin^2\theta d\varphi^2$ is 
the 
metric on the unit 2-sphere, the 
constant $H=\sqrt{\Lambda/3} $  is the 
Hubble parameter of the de Sitter background, $\Lambda>0$ 
is the cosmological constant and $m>0$ is a second 
parameter describing the mass of the central inhomogeneity 
({\em e.g.}, \cite{Bousso}). 
In general, the locations of the apparent horizons for a 
spherically-symmetric system can be calculated from the 
radial element of the inverse metric $g^{rr}=0$ 
\cite{hobson, nielsen}. Thus the apparent horizons for the 
Schwarzschild-de Sitter-Kottler solution are defined by the 
positive roots of the cubic equation
\be\label{3}
1-\frac{2m}{r} -H^2r^2 =0.\label{eq:SDSKhor} 
\ee
Following the method outlined by Nickalls in 
\cite{nickalls}, these roots may be written as
\begin{eqnarray}
r_1&=&\frac{2}{\sqrt{3}H}\sin\theta,\nonumber\\
r_2&=&\frac{1}{H}\cos\theta-\frac{1}{\sqrt{3}H}\sin\theta,\nonumber\\
r_3&=&-\frac{1}{H}\cos\theta-\frac{1}{\sqrt{3}H}\sin\theta,\label{eq:horizons}
\end{eqnarray}
where $\sin (3\theta)=3\sqrt{3} \, mH$.  Since $m$ and $H$ 
are 
both necessarily positive (we only consider expanding 
universes), $r_3$ is negative and therefore 
unphysical.  We thus refer to this spacetime as having only 
two apparent horizons.  We refer to $r_1$ as the black hole 
apparent horizon, since it reduces simply to the 
Schwarzschild horizon at $2m$ if there is no background 
expansion $H\rightarrow 0$, and we refer to $r_2$ as the 
cosmological apparent horizon, since it reduces to the 
static de Sitter horizon at $1/H$ if there is no mass 
present. 
The metric~(\ref{1}) is static in the 
region covered by the coordinates $\left( t, r, 
\theta, \varphi \right)$, which is comprised between these 
two horizons. 

A number of interesting observations can be made.  
First, both apparent horizons only actually exist if 
$ 0<\sin( 3\theta) < 1$. In this case, since the metric is 
static between these two horizons, the apparent black hole 
and  cosmological horizons are also event horizons and, 
therefore,  null  surfaces.  Second, if 
$\sin (3\theta)=1$  it is easy to show that these horizons 
then coincide.  This case corresponds to the 
Nariai black hole.  Finally, for $\sin (3\theta) >1$ both 
horizons become complex-valued and therefore unphysical, 
and one is left with a naked singularity. These results can be summarized as follows:
\begin{eqnarray}
mH<1/(3\sqrt{3})&\rightarrow&\text{2 horizons }r_1 {\text{ and }} r_2,\nonumber\\
mH=1/(3\sqrt{3})&\rightarrow&\text{1 horizon }r_1=r_2,\nonumber\\
mH>1/(3\sqrt{3})&\rightarrow&\text{no horizons }.
\end{eqnarray}
 
The Hubble parameter for an idealized de Sitter background 
is a constant, whereas more realistic models incorporate a 
time-dependent Hubble parameter.  With a clear 
understanding of the static horizons in the 
Schwarzschild-de Sitter-Kottler spacetime, we may now 
study the dynamical horizons which emerge by considering a 
more realistic time-dependent metric.

\section{Apparent horizons of the McVittie metric}

We now consider the McVittie metric for a black hole 
embedded in an FLRW background which is expanding with the 
Hubble flow \cite{McVittie}. For simplicity, we restrict 
ourselves to 
the case in which the background is spatially 
flat (curvature index $K=0$). The line element can thus be cast 
in the form \cite{Roshina}
\be
ds^2=-\left[ 1-\frac{2m}{r}-H^2(t) \right] dt^2 
-\frac{2H(t) r}{\sqrt{1-\frac{2m}{r}}} \, dtdr 
+r^2d\Omega_{(2)}^2 \,. \label{4}
\ee
Here $H(t) \equiv \dot{a}(t)/a(t)$, where $a(t)$ is the scale 
factor of the FLRW background and an overdot denotes 
differentiation with respect to the comoving time $t$.   
Note that for the case of a static background in which 
$a(t)=\exp(\sqrt{\Lambda/3} \, t)$ and 
$H=\sqrt{\Lambda/3}$, the McVittie metric  actually 
corresponds to the Schwarzschild-de 
Sitter-Kottler metric given by~(\ref{eq:SDSK}) via a 
simple transformation of the time coordinate 
\cite{arakida}.  Assuming a perfect fluid stress energy 
tensor, we may use Einstein's equations to calculate forms 
for the density $\rho(r,t)$ and pressure $P(r,t)$ of the 
background fluid in McVittie's metric. The density turns 
out to correspond to the known FLRW density
\be
\rho(t)=\frac{3}{8\pi} \, H^2(t) \,,
\ee
One may consider 
arbitrary FLRW backgrounds generated by cosmic fluids 
satisfying any equation of state (in fact, in the next 
section, we study a FLRW universe dominated by a phantom 
fluid). For illustrative purposes however, in this section we restrict our attention to a cosmic fluid which reduces 
to dust at spatial infinity.  This corresponds to an
equation of state parameter $w=0$, so the pressure can be shown to be \cite{Roshina}
\be
 P(t,r)=\rho(t) \left( 
\frac{1}{\sqrt{1-\frac{2m}{r}}} -1 \right) \label{5} \,.
\ee

Other quantities may be calculated from the inverse metric, given by
\be
\left( g^{\mu\nu} \right)=\left(
\begin{array}{cccc}
 -\, \frac{1}{1-2m/r} & -\, \frac{Hr}{\sqrt{1-2m/r}} & 0 & 
0 \\
&&& \\
-\, \frac{Hr}{ \sqrt{1-2m/r} } & \left(1-\frac{2m}{r} 
-H^2r^2  \right)   & 0 & 0 \\
&&&\\
0 & 0 & \frac{1}{r^2}  & 0 \\
&&&\\
0 & 0& 0 & \frac{1}{r^2 \sin^2 \theta} 
\end{array} \right)\,. \label{6} 
\ee
The Misner-Sharp-Hernandez mass $M_{MSH}$ 
\cite{MisnerSharpHernandez} contained in a sphere of areal 
radius $r$ is defined, in the case of  spherical symmetry, 
by
\be
1-\frac{2M_{MSH}}{r} =g^{rr}.
\ee
Thus, we obtain
\be 
M_{MSH}=\frac{4\pi G}{3} \, \rho \, r^3 +m \,.
\ee
which is interpreted as the contribution of the energy of 
the cosmic fluid contained  in the ball plus the 
contribution of the local inhomogeneity.  This mass 
coincides with the Hawking-Hayward quasi-local mass 
\cite{HawkingHayward}.

Since for the McVittie metric  $r$ is an areal radius and 
the system is spherically 
symmetric, the apparent horizons can once again be 
calculated from $g^{rr}=0$, corresponding to
\be\label{8}
1-\frac{2m}{r}-H^2(t) \, r^2=0 \,.
\ee
This is clearly equivalent to the Schwarzschild-de 
Sitter-Kottler horizon condition given by 
(\ref{eq:SDSKhor}) but with a time-dependent Hubble 
parameter.  We denote the resulting time-dependent apparent 
horizons $r_1(t)$ and $r_2(t)$, and these correspond to the 
solutions $r_1$ and $r_2$ given in equation 
(\ref{eq:horizons}) but with the replacement  $H\rightarrow 
H(t)$.  Since the apparent horizons for the  McVittie 
metric are dynamical, rather than static,  their relative 
locations now depend on the cosmic time.

\subsection{Dynamics of the apparent horizons}

Analogous to the Schwarzschild-de Sitter Kottler case, the 
condition for both horizons to exist is $0<\sin 
( 3\theta ) <1$, which corresponds to 
$mH(t)<1/(3\sqrt{3})$ (and $mH(t)>0$ which is always 
satisfied).  
However, unlike the former case where the Hubble parameter 
is a constant, this inequality will only be satisfied at 
certain times during the cosmological expansion, and not 
at others.  The time at which $mH(t)=1/(3\sqrt{3})$ is 
unique for a dust-dominated background with $H(t)=2/(3t)$, 
and we denote it $t_* =2\sqrt{3} \, m$.  The three cases 
may 
then be characterized as:

\begin{itemize}

\item $t<t_*$: at early times $m>\frac{1}{3\sqrt{3} 
\,H(t)}$, 
so both $r_1(t)$ and $r_2(t)$ are complex and therefore 
unphysical.  {\em There are no apparent horizons.}

\item $t=t_*$: at this time $m=\frac{1}{3\sqrt{3}\,H(t)}$ 
and the horizons $r_1(t)$ and $r_2(t)$ coincide at a real, 
physical location. {\em There is a single apparent horizon at $\frac{1}{\sqrt{3}\,H(t)}$}.
 
\item $t>t_*$: at late times $m<\frac{1}{3\sqrt{3}\,H(t)}$, 
so both $r_1(t)$ and $r_2(t)$ are real and therefore 
physical.  {\em There are two apparent horizons.}

\end{itemize}

\begin{figure}
\includegraphics{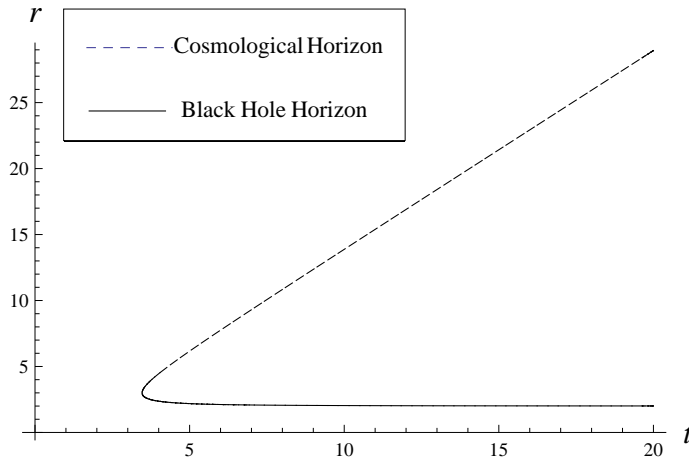}
\caption{the behaviour of the McVittie apparent  horizons 
versus time in a dust-dominated  background 
universe. We arbitrarily fix  $m=1$, hence time $t$ 
and  radius $r$ are measured in units of $m$ (see text for 
details).\label{figure1}}
\end{figure}

The qualitative dynamical picture which emerges from this 
analysis is  the following and is illustrated in 
fig.~\ref{figure1}.

The lack of apparent horizons for $t<t_{*}$ leaves a naked 
singularity at $r=2m$, where the Ricci scalar and pressure 
also diverge (see below).  This is explained by the 
divergence of the Hubble parameter $H(t)$ in the early 
universe, causing the mass $m$ to remain supercritical,  {\em i.e.}  causing $m>\frac{1}{3\sqrt{3}\,H(t)}$ to 
be 
satisfied.  Analogous to the Schwarzschild-de 
Sitter-Kottler solution,  a black hole horizon cannot 
be accommodated in such a small universe.

At the critical time $t_{*}$ a black hole apparent horizon 
appears and coincides with the cosmological apparent 
horizon at $r_1(t)=r_2(t)=\frac{1}{\sqrt{3}\,H(t)}$.  For a 
dust-dominated cosmological background this may be given as $r_1=r_2=3m$.  This is the analog of the Nariai black hole in the Schwarzschild-de Sitter-Kottler solution, but it is instantaneous. 

As time progresses, $t>t_{*}$, the single horizon splits 
into a dynamical black hole apparent horizon surrounded by 
a time-dependent cosmological horizon.  This solution can 
progressively constitute a better and better toy model for 
a spherical, non-accreting astrophysical black hole in the     
late universe with $mH \ll \frac{1}{3\sqrt{3}}\approx 
0.192$.   The black hole apparent horizon shrinks, 
asymptoting to the spacetime singularity at $2m$ from above as $t\rightarrow +\infty$, while the cosmological apparent horizon expands monotonically, tending to $1/H(t)$ in the same limit. 

The actual universe is of course not dust-dominated, and is  
better described by the scale factor for expansion 
\begin{equation}
a(t)=\left[\frac{(1-\Omega_{\Lambda,0})}{\Omega_{\Lambda,0}} 
\sinh^2\left(\frac{3}{2}H_0\sqrt{\Omega_{\Lambda,0}}t\right)\right]^{1/3},
\end{equation}
consistent with the spatially flat concordance model 
\cite{hobson}.  Here $H_0\approx70$ km s$^{-1}$ Mpc$^{-1}$ 
is the current value of the Hubble parameter and 
$\Omega_{\Lambda,0}\approx0.7$ is the current dark energy 
density.  Using this we may calculate actual values for 
$t_{*}$ and apparent horizon locations for black holes in 
our universe. Considering, for example, the $~  10^6 
M_{\odot}$  black hole at the centre of the Milky Way, we 
find that a  single horizon would have first appeared as 
early as $t_{*}\approx 17$  secs and at a radius very close 
to the centre $r_1(t_{*})= r_2(t_{*})\approx 
1.4\times10^{-7}$pc.  Thereafter, this would have split 
into two apparent horizons,  which would have become 
increasingly separated.  Note that a problem with this 
calculation is that it  neglects mass accretion. The 
results are therefore  purely theoretical and would only 
truly be valid if this  black hole had always existed at 
its current mass. Although in reality there were no bound 
structures in the universe at such an early time, this 
calculation does at least provide some insight into the 
scales involved.

Let us discuss now the well known singularity \cite{Nolan, 
Roshina, AbdelqaderLake}. The surface of equation 
$f(r) 
\equiv r-2m =0$ 
has normal 
$N_{\mu}=\nabla_{\mu}f=\delta_{1 \mu}$ with norm squared
\be
N_{\mu}N^{\mu}= g^{\mu\nu} N_{\mu}N_{\nu} \left|_{r=2m} 
\right.=-4m^2H^2(t) <0 \,.
\ee
$N^{\mu}$ is timelike and the surface $r=2m $ is 
spacelike. The Ricci scalar
\be
R=-8 \pi T^{\mu}_{\mu}=8\pi \left( \rho -3P \right)=8\pi 
\rho(t) \left( 4-\frac{3}{\sqrt{1-\frac{2m}{r} } } \right) 
\ee
diverges as $r\rightarrow 2m^{+}$. This singularity 
separates spacetime into two disconnected regions $r<2m$ 
and $r>2m$ \cite{Nolan}; the latter region is described by 
the metric~(\ref{4}). At the critical time $t_*$, when 
$r_1(t)=r_2(t)=1/ (\sqrt{3}H(t))$, the normal to the 
surface of equation ${\cal F}(r)   \equiv 
r-1/(\sqrt{3}H(t))=0$ is  $M_{\mu}=\nabla_{\mu}  {\cal 
F}=\delta_{1  \mu}$ and 
\be
M^{\mu}M_{\mu} = g^{11}\left(r=\frac{1}{\sqrt{3}H(t)}\right)=\frac{2}{3} \left( \frac{1}{3}-\sqrt{3}mH(t) 
\right) =0.
\ee
Thus the (cosmological and black hole) apparent horizon is 
instantaneously null. 

By differentiating the cubic equation~(\ref{8}), 
one may solve for the rate of change in location of 
the apparent horizons with respect to the comoving time.  
Dropping the $t$-dependencies for simplicity, one 
obtains 
\be
\dot{r}_{AH}=-\, \frac{2H\dot{H} r_{AH}^3 }{ 3H^2 r_{AH}^2 
-1}. 
\ee
Rearranging this, one can compare the expansion 
rates of the apparent horizons with that of the cosmic 
substratum,
\be
\frac{\dot{r}_{AH} }{r_{AH} }-H= -H\left( 1+ 
\frac{2\dot{H}r_{AH}^2 }{3H^2 r_{AH}^2 -1} \right) \,.\label{eq:rateofexp}
\ee
This equation shows that the apparent horizons are not 
comoving except for trivial cases. This explains why  the 
black  hole cannot remain static but is instead  forced to 
expand \footnote{Similarly, wormholes embedded in 
cosmological backgrounds must expand \cite{wormholes}.}. In 
the case of  a spatially flat FLRW universe (without the 
central inhomogeneity), it turns out that even the single 
cosmological horizon  at $r_{AH}(t) \equiv r_c(t)=1/H(t)$ 
is not 
comoving, since
\be
\frac{ \dot{r}_c }{ r_c }=-\frac{ \dot{H} }{ H } \neq H \,.
\ee


\subsubsection{Horizon entropy} 

\begin{figure}
\includegraphics{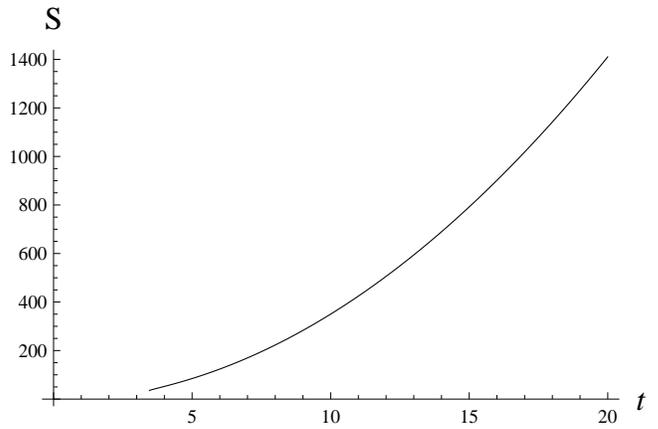}
\caption{\label{figure2}  the total horizon entropy 
$S$ (in units $\frac{k_Bc^3}{\hbar G}$, where $k_B$ 
is the  Boltzmann constant) associated  with the apparent 
horizons as a function of time.}
\end{figure}

It is widely believed that, in the absence of event 
horizons, an entropy can be meaningfully ascribed to 
apparent horizons. The thermodynamics of these horizons 
has been discussed extensively \cite{AHthermodynamics}. 
Therefore, it is interesting to ask whether the total 
entropy associated with both the black hole and 
cosmological 
apparent horizons is a non-decreasing function of time. 
The area $A_1$ of the black hole apparent horizon is 
decreasing, but it is bounded from below while this 
behaviour is more than compensated for by the 
increase of the area $A_2$ of the cosmological apparent 
horizon. The total horizon entropy
\be
S=S_1+S_2 = \pi \left( r_1^2 +r_2^2 \right)
=\frac{A}{4} \,, 
\ee
where $A=A_1+A_2$, is plotted in fig.~\ref{figure2}.

Since the apparent horizons emerge as a pair at $t=t_*$, 
the horizon entropy $S$ exhibits a discontinuous jump from 
zero value at this time.

\subsection{A phantom background}

We now discuss the situation of a cosmological background 
dominated by a phantom fluid with equation of state
satisfying $P+\rho<0$ ($w=P/\rho<-1$) and violating the weak energy 
condition. 
The recent renewed interest in such a field has been  
motivated by the analysis of data from supernovae 
Ia \cite{SN} and the study of the effects of the 
accelerating  universe 
\cite{AmendolaTsujikawabook}.  The consideration of a 
phantom background has also led to the prediction  of a Big 
Rip singularity at a finite time in the  future 
$t_{\text{rip}}$ \cite{Caldwell}. We now 
consider a phantom background in the context of the 
McVittie solution.  Surprisingly,  this is a situation 
which has not received much attention in previous studies.

\begin{figure}
\includegraphics{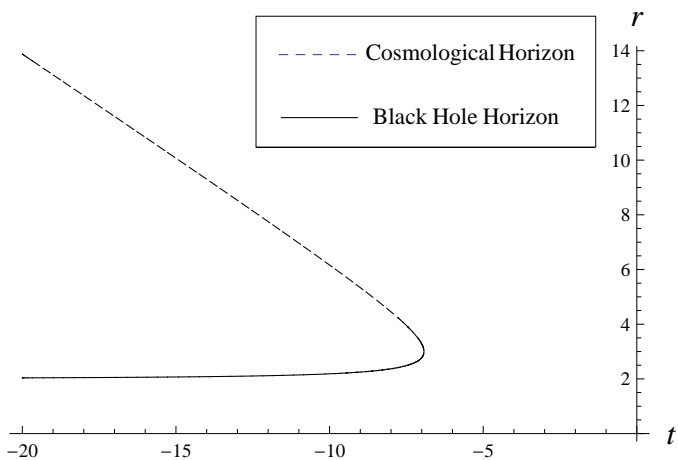}
\caption{\label{figure3} the behaviour of the 
McVittie apparent  horizons versus time in a 
phantom-dominated background universe for the parameter 
values $w=-1.5$ and $t_{\text{rip}}=0$.
}
\end{figure}

One may consider the late time 
behaviour of the Friedmann equation governing a phantom 
fluid and solve it to obtain a form for  the scale factor 
in terms of $  t_{\text{rip}}$ and $w<-1$.  Indeed the 
solution has been shown to be 
\cite{Caldwell}: 
\be
a(t)=\frac{A}{ \left( t_{\text{rip}}-t 
\right)^{\frac{2}{3| w+1|} }}, 
\ee
where $A$ is a constant. The Hubble parameter may 
therefore be written concisely as
\be
H(t)=\frac{2}{3|w+1|} \, \frac{1}{t_{\text{rip}}-t} \,.
\ee
Note the reverse behaviour of this function compared with 
the Hubble parameter for a dust-dominated universe 
$H(t)=2/(3t)$.  The latter diverges at the big bang 
singularity and gradually decreases over time, tending to 
zero.  The Hubble parameter for a phantom fluid, however, 
takes on a finite value at $t=0$ and slowly increases 
until the Big Rip time, at which point it too diverges.  
This suggests that the horizons around black holes 
embedded in a phantom fluid might behave in the opposite 
way to those in a dust-dominated background with $w>-1$.  
Indeed this does turn out to be the case, and the 
discussion in the previous subsection can be repeated. 
The result is plotted in fig.~\ref{figure3}.

We may summarize our results in an expanding universe 
dominated by a phantom fluid as follows. In the early 
universe, both black hole and 
cosmological apparent horizons exist,  and are 
approximately located at $2m$ and $1/H(t)$, respectively.  
As time progresses  the cosmological horizon shrinks and 
the black hole  horizon expands, until they meet and merge 
at the  critical time $t_*$.  Thereafter they disappear,  
leaving 
behind a naked 
singularity. During this evolution the total 
apparent horizon entropy decreases 
and has a discontinuous jump to zero value at $t_*$. This 
behaviour is yet another manifestation of the 
``weirdness'' of the phantom fluid, which seems to violate 
the second law of thermodynamics in many ways 
\cite{phantomthermodynamics}. 

The behaviour of the apparent horizons for a phantom cosmic 
background was derived in Ref.~\cite{Gaoetal1} for {\em 
generalized} McVittie solutions, which are obtained by 
relaxing the McVittie no-accretion condition and allowing 
for a radial energy flux onto the black hole 
\cite{FaraoniJacques, Gaoetal1}. For simplicity of 
modelling, this radial flux density $q^{\mu}$  is 
necessarily  
spacelike and violates the energy conditions. The lesson to 
be learnt by the present discussion of the corresponding 
McVittie solution  with $q^{\mu} \equiv 0$ is that the 
disappearance of the apparent horizons is not due to the 
fact that the accreted phantom fluid violates the 
weak energy condition and the total accreted mass becomes 
zero: it is due to the phantom character of the fluid which 
dictates the unusual cosmic expansion leading to the Big 
Rip, but not to accretion.

\section{Discussion and conclusions}

In order to understand the bizarre phenomenology of 
apparent horizons in the McVittie spacetime, it is useful 
to first understand the Schwarzschild-de Sitter-Kottler 
solution of the Einstein equations. This is a special case 
of the McVittie solution.   Our study of the simple, 
static, Schwarzschild-de Sitter-Kottler metric  has 
essentially revealed that a black hole can only fit 
in a de Sitter universe if its horizon size (determined by 
its mass) does not exceed the size of the cosmological 
horizon. Equipped with this clarity, we have then moved  
on to consider the  more complicated McVittie solution, 
which accounts for a  dynamical background and thus better 
represents reality.  Not surprisingly, the condition for 
the existence of the apparent horizons in this case is 
analogous to the corresponding one in the static case, with 
the static Hubble  constant replaced by a dynamical Hubble 
parameter.  This follows  from the dynamical nature of the 
apparent horizons themselves  in this case, which we are 
able to locate throughout their period of existence.  The 
absence 
of any (black hole or cosmological) apparent horizons at 
early times is now easily  understood. At early 
times the mathematical solutions suggest that the 
cosmological horizon would be smaller than  the black hole 
horizon, but this is not possible since the  universe at 
this time would be too small to accommodate  a black hole 
apparent horizon at all. One cannot then meaningfully 
distinguish between the ``black hole'' 
and the ``universe'' in which it is embedded; rather, the 
mathematical solutions represent neither and do not possess 
the properties of a black hole or a universe.   
Thus at early times, not only is there a naked 
singularity, but 
the cosmological apparent horizon is also absent. The presence 
of this naked singularity prevents one from being able  to 
derive the 
McVittie solution as the development of regular Cauchy 
data.  At some finite time, given by $3m$ for a  
dust-dominated background, the cosmological solution is 
able to catch up with the black hole solution and a single 
black hole/cosmological apparent horizon appears.  These then split 
and continue to diverge thereafter.

The McVittie metric does not account for accretion onto 
the central mass.  Hence the mass parameter $m$ is fixed 
and the horizon dynamics are wholly determined by the 
expansion of the universe.  If the no-accretion  
assumption is relaxed however, the black hole mass itself 
is then also  determined by the universe's expansion 
(possibly with some residual 
freedom) and cannot be fixed {\em a priori}. Indeed some 
generalized McVittie solutions, for which $m$ becomes  
a function of time, have already been derived  
\cite{FaraoniJacques, Gaoetal1}.  At late times 
this class of solutions converges to an attractor with a 
well-defined  mass function $m(t)$ \cite{Gaoetal2}. 
Other solutions for cosmological 
lumps (including black holes) have also been  derived and 
investigated without  imposing the no-accretion condition 
in general relativity  and in  scalar-tensor and higher 
derivative gravity  \cite{cosmologicalblackholes, Barrow, 
CliftonMotaBarrow, Clifton}. In  some of these 
studies, the  phenomenology of the apparent 
horizons appears to be even more bizarre than in the  
McVittie case and involves the creation or 
disappearance also of inner black hole apparent horizons 
\cite{HusainMartinezNunez, myClifton, 
FaraoniVitaglianoSotiriouLiberati}. 

Locating the apparent horizons and understanding, at least 
in principle, their behaviour is not the whole story. The recent work \cite{AbdelqaderLake} studying the 
global structure of the McVittie solution has unveiled a 
new  feature which is believed to be generic: radial 
ingoing null geodesics do not penetrate the black hole 
apparent horizon to reach the $r=2m$ singularity, but are 
asymptotic to this horizon. In our opinion, this feature is 
not too surprising for a solution in which radial flow onto 
the central black hole is excluded by construction. The 
property of radial ingoing null geodesics merely 
reflects the McVittie no-accretion condition. In fact, the 
ingoing radial null geodesics can be seen as the 
test-particle limit of a gravitating null dust (which 
however, would be forbidden by the no-accretion 
condition and could not fit in the McVittie spacetime). 
Future work  to fully understand this feature, as well as 
more general solutions representing black holes embedded in 
cosmological backgrounds, will be presented elsewhere.

\begin{acknowledgments}

VF thanks Bishop's University and the Natural  Sciences and 
Engineering  Research Council of Canada ({\em NSERC}) for 
financial support. 

\end{acknowledgments}



\end{document}